\documentclass[journal=nalefd,manuscript=article]{achemso}

\usepackage[version=3]{mhchem} % Formula subscripts using \ce{}

\author{Marcos H. D. Guimar\~{a}es}
\email{m.h.diniz.guimaraes@rug.nl}
\author{A. Veligura}
\author{P. J. Zomer}
\author{T. Maassen}
\author{I. J. Vera-Marun}
\author{N. Tombros}
\author{B. J. van Wees}
\affiliation[University of Groningen]
{Physics of Nanodevices, Zernike Institute for Advanced Materials, University of Groningen, The Netherlands}

\title
{Spin transport in high quality suspended graphene devices}

\begin{document}
%%%%%%%%%%%%%%%%%%%%%%%%%%%%%%%%%%%%%%%%%%%%%%%%%%%%%%%%%%%%%%%%%%%%%
%% The manuscript does not need to include \maketitle, which is
%% executed automatically.  The document should begin with an
%% abstract, if appropriate.  If one is given and should not be, the
%% contents will be gobbled.
%%%%%%%%%%%%%%%%%%%%%%%%%%%%%%%%%%%%%%%%%%%%%%%%%%%%%%%%%%%%%%%%%%%%%
\begin{abstract}
We measure spin transport in high mobility suspended graphene ($\mu \approx 10^{5}$cm$^{2}$/Vs), obtaining a (spin) diffusion coefficient of 0.1 m$^{2}$/s and giving a lower bound on the spin relaxation time ($\tau_{s}\approx$ 150 ps) and spin relaxation length ($\lambda_{s}$=4.7 $\mu$m) for intrinsic graphene.
We develop a theoretical model considering the different graphene regions of our devices that explains our experimental data.
\end{abstract}

The prospectives for graphene spintronics are very positive, with theoretical predictions of spin relaxation times ($\tau_{s}$) in the order of hundreds of nanoseconds and higher\cite{EPJ-Brataas2007,PRL-Brataas2009} due to the lack of nuclear spin for carbon's most common isotope ($^{12}$C) and weak intrinsic spin-orbit coupling.
% and exceptional electronic properties.
On the other hand the experimental results typically find $\tau_{s}$ of hundreds of picoseconds \cite{PRB-Csaba2009,PRL-Kawakami2010,Nanoletters-Barbaros2011}.
In order to clarify the limitations and mechanisms for spin relaxation in monolayer graphene devices a lot of effort has been done by both experimentalists \cite{Nature-Niko2007,PRB-Mihai2009,PRB-Csaba2009,PRL-Kawakami2010,PRL-Kawakami2011,Nanoletters-Barbaros2011,PRB-Thomas2011} and theoreticians \cite{PRL-Castroneto2009,PRL-Brataas2009,PRL-Guinea2012,NJP-Wu2012} but up to now this topic is still under debate.
Theoretical results pointed out that the D'Yakonov-Perel mechanism \cite{SJETP-Dyakonov1971} and locally enhanced spin-orbit coupling due to impurities and lattice deformations \cite{PRL-Brataas2009,PRL-Castroneto2009} are the most probable mechanisms for spin relaxation for measurements in graphene spintronic devices \cite{PRL-Brataas2009,PRL-Guinea2012,NJP-Wu2012}.
Since all devices studied so far were fabricated on substrates (SiO$_{2}$ \cite{Nature-Niko2007,PRB-Mihai2009,PRB-Csaba2009,PRL-Kawakami2010,PRL-Kawakami2011,Nanoletters-Barbaros2011} or SiC\cite{Nanolett-Thomas2012}) in which the intrinsic properties of graphene are masked due to the electronically coupling to the substrate\cite{Natphys-Crommie2009}, no previous study was able to confirm these recent theoretical predictions.
In order to address this problem we study spin transport in high-mobility suspended graphene devices.
By removing the substrate, thereby suspending the graphene flake, we are not only capable of achieving a high quality device with low contamination \cite{SSC-Kim2008,Natnano-Andrei2008}, but we can also investigate how the absence of the rough SiO$_{2}$ substrate influences the spin transport in graphene.
Moreover it opens the possibility to exploit the exquisite mechanical properties of graphene \cite{Nanolett-Mceuen2008} and to study how pseudo-magnetic fields and strain \cite{Natphys-Geim2010,Science-Crommie2010} affect spins in graphene, paving the way for graphene in the field of spin-nanomechanical applications \cite{NatNano-Mohanty2008,PRL-Stiles2010,PRL-Kovalev2011}.

The effective spin-orbit (SO) field that the spins experience in graphene is directly related to the density of impurities and adatoms \cite{PRL-Castroneto2009,NJP-Wu2012}.
This effective SO field causes a momentum-dependent spin precession, resulting in spin decoherence and relaxation.
This mechanism is known as the D'Yakonov-Perel mechanism for spin relaxation \cite{SJETP-Dyakonov1971}.
Assuming such a mechanism for spin relaxation\bibnote{Up to now all previous results for graphene spin valves on SiO$_{2}$ or for our suspended devices without cleaning show a constant behaviour of $\frac{\tau_{s}}{\tau_{p}}$ as a function of the Fermi energy. This goes against the theories for Elliott-Yafet mechanism for spin relaxation in graphene as showed in Ref. 11. This way we assume the most recent theory that best describe all the measurements of spin transport in graphene available so-far (Ref. 12).} with an effective SO coupling $\Delta_{SO}$, the spin relaxation time behaves like \cite{SJETP-Dyakonov1971,NJP-Wu2012}:

\begin{equation}
\label{eq:DP}
\frac{1}{\tau_{s}}=\frac{4 \Delta_{SO}^{2}}{\hbar^{2}}\tau_{p}
\end{equation}

\noindent where $\tau_{p}$ is the momentum relaxation time and $\hbar$ the reduced Planck constant.
With the reduction of adatoms and impurities we should obtain a low value of $\Delta_{SO}$ approaching the theoretical predictions\cite{PRL-Brataas2009,PRB-MacDonald2006,PRB-Fabian2009} ($\Delta_{SO}^{theory}\approx$ 10 $\mu$eV) and a high value for the momentum relaxation time $\tau_{p}>0.1$ ps.
%For clean (high mobility) samples we expect a reduction in $\Delta_{SO}$ and an increase in $\tau_{p}$.
Since $\tau_{s}$ increases quadratically with the reduction in $\Delta_{SO}$, we expect that for clean samples we reach the initial theoretical limits of $\tau_{s}$ > 10 ns.

At a first glance, suspending the graphene flake to obtain a very high quality device seems the best approach.
But the main challenge is that the most common technique to suspend graphene flakes \cite{SSC-Kim2008,Natnano-Andrei2008} is acid-based and therefore not compatible with ferromagnetic metals necessary for spin transport.
The reason is that the acid used to etch the SiO$_{2}$ underneath the graphene also etches away the ferromagnetic metals used for the electrodes.
Also, devices produced by the standard technique are typically short, in the order of 1 $\mu$m or less to avoid that the graphene flakes collapses when a gate-voltage is applied.
This is undesirable for spin precession measurements since the time for one period of precession for low magnetic fields ($B<1$ T) have to be in the same order as the diffusion time of the spins in order to obtain a Hanle precession curve \cite{APS-Fabian2007} showing a complete spin precession.
To overcome these issues we developed a polymer-based method \cite{JAP-Niko2011} in which we are able to produce flakes suspended over long distances and contacted by ferromagnetic electrodes with highly resistive barriers.
In our process the graphene flakes are exfoliated on top of a 1 $\mu$m thick Lift-Off Resist (LOR) film spin-coated on a Si/SiO$_{2}$ (500 nm) substrate.
Single layer graphene flakes to be used in our devices are selected by optical contrast \cite{APL-Geim2007,APL-Falko2007}.
The highly doped Si substrate is used to apply a back-gate voltage V$_{g}$ that induces a charge carrier density according to $n=\alpha (V_{g}-V_{0})$, where $\alpha=0.45 \times 10^{10}$ cm$^{-2}$V$^{-1}$ is the effective gate capacitance and $V_{0}$ is the position in gate voltage of the minimum of conductance.
Using standard electron beam lithography (EBL) and metal evaporation methods we deposit the 1 nm thick Al$_{2}$O$_{3}$ tunnel barriers and the 60 nm Co contacts (Figure 1a).
A second EBL step is used to make parts of the graphene flake suspended (Figure 1 b).
The detailed device fabrication is described in the supporting information.
The resistance of our contacts (usually R$_{C}\sim$20 k$\Omega$) and the gate-voltage dependence of our devices are characterized at the beginning of every set of measurements.
Before the quality improvement via the current annealing procedure (described in the next paragraph), the devices typically show high p-doping and only a small change of the resistance as a function of V$_{g}$ is observed \cite{JAP-Niko2011, NatPhys-Niko2011}.

To improve the quality of our suspended graphene devices we perform current annealing \cite{APL-Bachtold2007} in vacuum (at a pressure better than 10$^{-6}$ mbar) at a base temperature of 4.2 K (see supporting information).
Since the contact resistance of the spin injector and detector in our samples has to be kept at high values (>10 k$\Omega$) to avoid the impedance mismatch and contact induced spin relaxation \cite{PRB-Wees2000,PRB-Mihai2009,PRL-Kawakami2010}, we use the outer electrodes to apply the high current densities capable of heating the graphene flake to high temperatures (Figure 1c).
As shown in Figure 1b and c, our devices are composed of two supported outer parts with a central suspended part.
During the annealing the graphene flake is cooled by the contacts and the substrate in the supported parts and mainly heats up in the suspended region.
When a sufficiently high temperature is achieved, the polymer residues and contaminants are removed from the suspended central region.
Finally the high quality of the devices can be verified via the behaviour of the square resistance of the central region (R$_{sq}$) as a function of the gate-voltage (Figure 1e).
It is worth noting that the value and the gate dependence of the sheet resistance of the supported regions do not change significantly after the current annealing procedure. This means that the values of carrier density presented here are representative for the central suspended region.
We obtained two high quality devices in different flakes with mobilities higher than $\mu$ = 10$^{5}$ cm$^{2}$/Vs. One of the devices was current annealed twice and showed a mobility of $\mu \approx$ 1.2 $\times$ 10$^{5}$ cm$^{2}$/Vs after the first current annealing and $\mu \approx$ 3 $\times$ 10$^{5}$ cm$^{2}$/Vs after the second.
This device was also characterized for charge and spin transport in between the two current annealing steps.
Here we show the results for this representative high mobility sample after the second current annealing.
The separation between the inner electrodes (suspended region) is L = 2.5 $\mu$m and the electronic mobility is $\mu\approx$ 3 $\times$ 10$^{5}$ cm$^{2}$/Vs (Figure 1e).
Similar results for both charge and spin transport were obtained in the other devices.

The values for mobility were obtained in two different ways: either by using the equation $\mu = \frac{\sigma}{ne}$ and taking the value of $\mu$ at a carrier density of n = 1 x 10$^{10}$ cm$^{-2}$, or by fitting the conductivity curves with the formula $1/\sigma = 1/(ne\mu + \sigma_{0}) + \rho_{s}$ \cite{PRL-Sarma2007,APL-Paul2011}, where $\sigma$ is the graphene conductivity, $e$ the elementary charge, $\sigma_{0}$ the conductivity at the charge neutrality point and $\rho_{s}$ is the contribution of short range scattering.
The values obtained by both ways are consistent with each other and are in the order of 10$^{5}$ cm$^{2}$/Vs.
We often observe a background resistance in our $R_{sq}(V_{g})$ curves (Figure 1e).
The most probable source for this background resistance is the non-cleaned region underneath and/or close by the contacts.
We did not subtract this background resistance in our calculations since this would only lead to higher mobility values.
%The specific values for each sample can be find in the supporting information.

\begin{figure}
	\centering
		\includegraphics[width=1.00\textwidth]{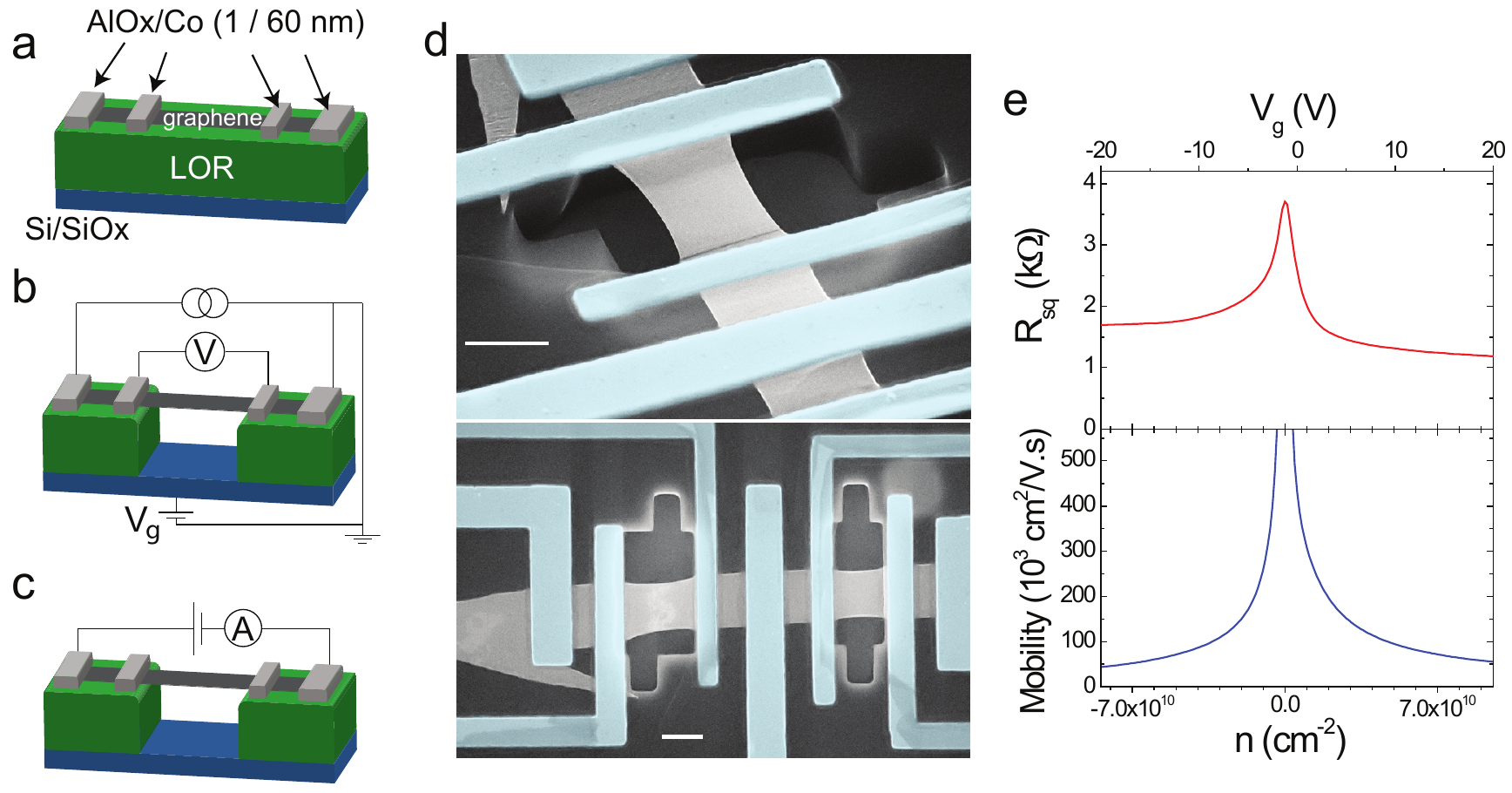}
	\caption{(a) The graphene flakes are exfoliated on a LOR film spin-coated on a Si/SiO$_{2}$ substrate followed by the deposition of Co contacts with aluminium oxide insulating barriers to avoid the conductivity mismatch. (b) Local 4-probe measurement geometry to characterize the graphene resistance as a function of the gate-voltage applied to the highly doped Si substrate. A second electron beam lithography step is performed to suspend the central region. (c) Scheme of the current annealing setup used to clean the graphene devices. The high DC bias is applied to the outer electrodes to avoid the degradation of the inner contacts since these are to be used for spin injection/detection. (d) Scanning electron micrographs of a typical device. The scale bars are 1 $\mu$m. (e) Local 4-probe resistance (top) and mobility (bottom) versus carrier density and gate-voltage after the current annealing procedure at 4.2 K. The carrier density shown corresponds to the carrier density in the central suspended region (see main text).}
	\label{fig:figure-1}
\end{figure}

To perform the spin-transport measurements we use a non-local geometry where the charge current path is separated from the voltage contacts to exclude spurious signals \cite{Nature-Niko2007} (Figure 2a).
When we sweep the in-plane magnetic field we can align the two inner ferromagnetic contacts in a parallel or anti-parallel configuration.
To extract the spin diffusion coefficient D$_{s}$ and spin relaxation time $\tau_{s}$, we perform Hanle precession measurements and use the solutions for the Bloch equations in the diffusive regime \cite{APS-Fabian2007,PRB-Mihai2009}.
The Hanle precession measurements are done by measuring the non-local resistance as a function of an applied perpendicular magnetic field.
To eliminate background signals in our analysis we fit the data for the total spin-signal given by: $R_{s} = \frac{R_{nl}^{\uparrow\uparrow} - R_{nl}^{\uparrow\downarrow}}{2}$, where $R_{nl}^{\uparrow\uparrow(\uparrow\downarrow)}$ is the non-local resistance in the parallel (anti-parallel) configuration of the inner electrodes (Figure 2b).
The values for D$_{s}$ and $\tau_{s}$ obtained for the suspended non-annealed samples are typically D$_{s}$ $\approx$ 0.02 m$^{2}$/s and $\tau_{s}$ $\approx$ 200 ps, giving a spin relaxation length of $\lambda_{s} = \sqrt{D_{s}\tau_{s}}$ $\approx$ 2 $\mu$m, showing no dependence on the temperature (from room temperature down to 4.2 K).
Since the results for the non-annealed samples show no substantial difference from measurements done in fully LOR supported devices (see supporting information), we can conclude that the substrate is not the main factor limiting the spin relaxation for our samples before the cleaning procedure.
When we consider roughness effects, this invariance of the spin relaxation time in our measurements without the rough substrate is in agreement with the results by Avsar et al. \cite{Nanoletters-Barbaros2011}, where it is shown that ripples in the graphene flakes have minor (or no) effects on the spin transport parameters.
Also, the widths of our contacts in our samples are much smaller than the spin relaxation length. So we believe that the regions underneath the contacts do not change significantly the spin transport properties in our measurements.

\begin{figure}
	\centering
		\includegraphics[width=1.00\textwidth]{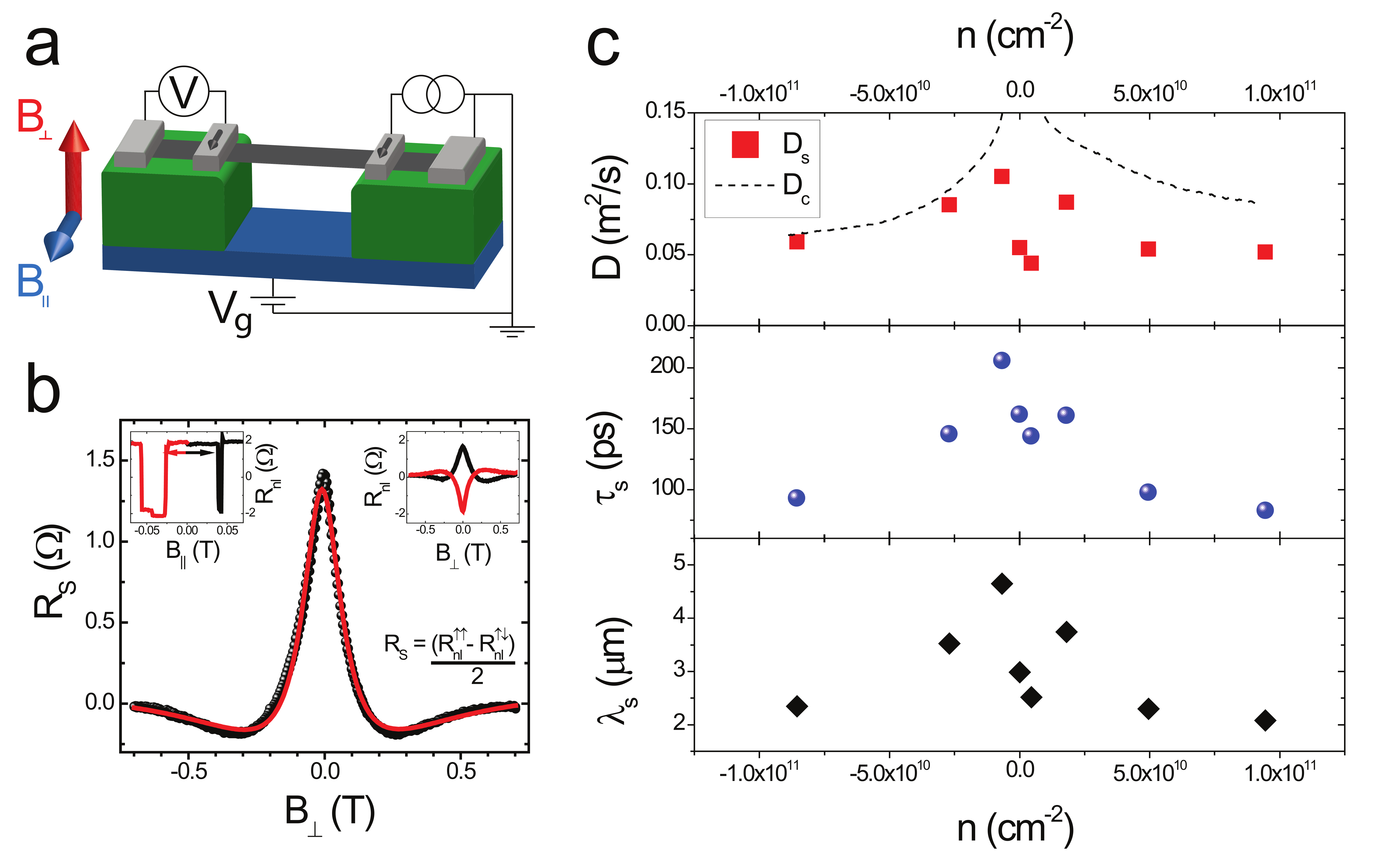}
	\caption{(a) The 4-probe non-local device configuration for Hanle precession and spin-valve measurements. For the precession measurements the two inner contacts are aligned in a parallel (anti-parallel) configuration and the non-local resistance is measured as a function of a perpendicular magnetic field B$_{\bot}$. (b) A measured Hanle precession curve (black circles) with a fit using the solutions for the Bloch equations (red line). The spin signal R$_{s}$ is recorded as a function of the perpendicular magnetic field. Left inset: the non-local resistance as a function of the parallel magnetic field. The positive (negative) values of $R_{nl}$ are due to parallel (anti-parallel) alignment of the inner electrodes. Right inset: the non-local resistance as a function of the perpendicular magnetic field for the parallel (black) and anti-parallel (red) configuration of the contacts. (c) Spin and charge diffusion coefficients D$_{s}$ (red squares) and D$_{c}$ (dashed line) respectively, spin relaxation time $\tau_{s}$ and spin relaxation length $\lambda_{s}$ as a function of the carrier density in the central suspended region. All the measurements shown were performed at 4.2 K.}
	\label{fig:figure-2}
\end{figure}

After cleaning by current annealing, we extract $\tau_{s}$ and $D_{s}$ as a function of the carrier density via Hanle precession measurements (Figure 2 c).
Comparing our results for the clean high-mobility samples to previous studies on SiO$_{2}$ supported devices\cite{Nature-Niko2007,PRB-Mihai2009,PRB-Csaba2009,PRL-Kawakami2010,PRL-Kawakami2011,Nanoletters-Barbaros2011} we observe an approximately 3 times higher spin diffusion coefficient while the values for the spin relaxation time remain similar.
Note that the experimental conditions allow us to extract a lower bound for the spin relaxation time in the suspended graphene region, this leads to important conclusions as we will discuss below. All the spin transport measurements presented here were performed at low temperatures (4.2 K).

We start by discussing the results for the spin diffusion coefficient.
In Figure 2c we have a comparison of the values of D$_{s}$ extracted from the Hanle precession measurements (red squares) with the value of the charge diffusion coefficient (dashed line) given by the Einstein relation: $D_{c}=1/(R_{sq}e^{2} \nu(E))$, where $\nu(E)$ is the density of states of graphene at the energy $E$.
In this case the square resistance, $R_{sq}$, was extracted by local 4-probe measurements, as depicted in Figure 1b.
It can be seen that D$_{s}$ and D$_{c}$ are in reasonable agreement, which is in accordance to previous works on exfoliated monolayer graphene \cite{PRB-Csaba2009}.

We now turn our attention to the results for the spin relaxation time $\tau_{s}$ (Figure 2c).
If the spin relaxation time in graphene is limited by a locally enhanced SO coupling \cite{PRL-Guinea2012,NJP-Wu2012} due to impurities and adatoms, we would expect a high $\tau_{s}$ for a high mobility sample, since an increased mobility is related to a reduction in the density of impurities.
Intriguingly, in our results for high mobility graphene spin-valves, in which the mean free path is in the order of a micrometer, $\tau_{s}$ is still in the order of hundreds of ps.
When we calculate the spin relaxation length $\lambda = \sqrt{D_{s} \tau_{s}}$ we obtain large values, up to $\lambda$ = 4.7 $\mu$m, even with the low spin relaxation time in our samples.
But one question remains: what limits the measured spin relaxation time in our high-quality devices?
To address this issue we performed numerical simulations of spin transport in our devices.

To properly represent our devices we extend the model adopted by Popinciuc et al. \cite{PRB-Mihai2009}.
We consider our devices as a three-part system separated by two boundaries. The central part represents the suspended region and the two identical outer parts represent the supported regions (Figure 3a).
The spins are injected at the left boundary by an injector of polarization $P_{c1}$ and contact resistance $R_{c1}$ and detected by a contact at the right boundary with polarization and contact resistance $P_{c2}$ and $R_{c2}$ respectively.
The spin diffusion coefficients, spin relaxation times and the conductivities, denoted by $D_{i(o)}$, $\tau_{i(o)}$ and $\sigma_{i(o)}$ respectively, for the inner (outer) parts can be defined independently.
We assume that the spin accumulation $\vec{\mu}_{s}(x)$ obeys the Bloch equation for diffusion in one-dimension with an applied magnetic field $\vec{B}$: $D_{s} \nabla^{2} \vec{\mu_{s}} - \frac{\vec{\mu_{s}}}{\tau_{s}} + \gamma \vec{B} \times \vec{\mu_{s}} = 0$, where $\gamma$ is the gyromagnetic ratio with $\gamma=g \mu_{B} \hbar^{-1}$ where $g$ the g-factor, $\mu_{B}$ the Bohr magneton and $\hbar$ the reduced Planck constant.
For the boundary conditions we assume that $\vec{\mu}_{s}$ is continuous and goes to zero at $x=\pm \infty$, the spin current at the right boundary is continuous and at the left boundary it is discontinuous by a value determined by the spin injection.
We also include the contact induced spin relaxation due to back-diffusion \cite{PRB-Mihai2009}, although for contact resistances in the order of those we encounter in our experiments, our simulations showed no substantial difference from the results considering the limit of infinite contact resistances.
Details of the model and simulation are included in the supporting information.

\begin{figure}
	\centering
		\includegraphics[width=0.50\textwidth]{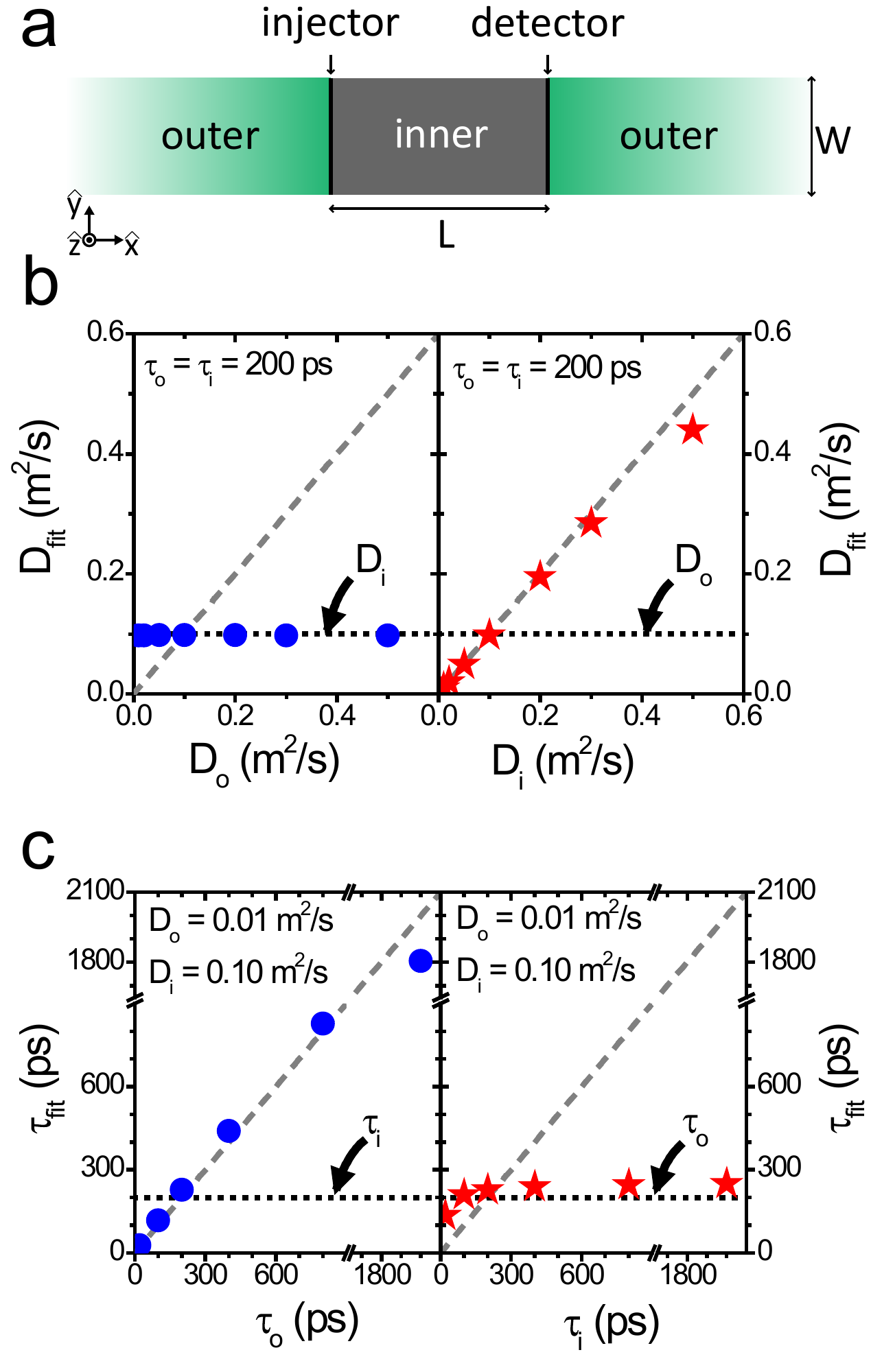}
	\caption{(a) A cartoon of the system used in our simulations: two identical and semi-infinite parts connected by one inner region with length L, all of width W. The spins are injected at the left boundary (x=0) and detected at the right boundary (x=L). (b) A graph showing the results obtained for the diffusion constant D$_{fit}$ by fitting the simulated data with D$_{i}$ (D$_{o}$) constant at the value represented by the dotted black line and changing D$_{o}$ (D$_{i}$) following the dashed grey line in blue circles (red stars). (c) Results for the spin relaxation time $\tau_{fit}$ by fitting the simulation data with $\tau_{i}$ ($\tau_{o}$) constant at the value represented by the dotted black line and changing $\tau_{o}$ ($\tau_{i}$) following the dashed gray line in blue circles (red stars). The resistivities of the tree regions were kept at 1 k$\Omega$ for the results in (b) and (c).}
	\label{fig:figure-3}
\end{figure}

By adding a perpendicular magnetic field $\vec{B}=B \hat{z}$ the spins can precess and the voltage at the detector electrode is calculated as a function of $B$, simulating the Hanle precession measurements.
The obtained data is then fitted the same way we fit our experimental data, with the solutions to the Bloch equation for a homogeneous system, and effective values for D$_{s}$ and $\tau_{s}$ are obtained and now denoted as D$_{fit}$ and $\tau_{fit}$ respectively.
These effective values are then compared to the values used in the simulation.

Since we have four different parameters to consider ($\tau_{i}$, $\tau_{o}$, $D_{i}$ and $D_{o}$), we change them one by one, keeping the others at a constant value.
We keep the sheet resistance of the three regions at the same value of 1 k$\Omega$.
The effect of varying the sheet resistance of the inner and outer regions are presented in the supporting information.
%Here we show the results for the diffusion coefficient and spin relaxation time.
Keeping $\tau_{i}=\tau_{o}=200$ ps and $D_{i}=0.1$ m$^{2}$/s, we varied $D_{o}$ from 0.01 to 0.5 m$^{2}$/s and extract $D_{fit}$ (Figure 3b, blue circles) which we compare to $D_{i}$ (dotted black line) and $D_{o}$ (dashed grey line).
It can be seen that a change in the spin-diffusion coefficient of the outer parts does not influence the results obtained by the Hanle precession fits in the studied range.
For all values of $D_{o}$ we always obtained a value close to the value of $D_{i}$, $D_{fit}\approx$0.1 m$^{2}$/s.
Keeping now $D_{o}$ constant and varying $D_{i}$ (Figure 3b, red stars) we confirm that from the Hanle precession analysis we always obtain a value $D_{fit}$ very close to the actual value of $D_{i}$ (dashed grey line).
Such a result agrees with our experimental measurements, where the values obtained for $D_{s}$ from the Hanle precession measurements are in good agreement with the charge diffusion coefficient, $D_{c}$, for the inner suspended part.
In other words, the diffusion coefficient we measure for our inhomogeneous devices is determined by the one in the central high mobility region.

Having confirmed that the results obtained for the diffusion constant from our experiments and simulations follow the same trend, we now analyze the effect of the spin relaxation times.
We fix the spin diffusion coefficients at $D_{i}=0.1$ $m^{2}/s$ and $D_{o}=0.01$ $m^{2}/s$, which are approximately the values we find in our experiments, and keep $\tau_{i}=200$ ps constant (dotted black line).
We then vary $\tau_{o}$ from 20 to 2000 ps (Figure 3c, blue circles).
The values obtained from the fit of the simulated data for $\tau_{fit}$ follow the increase in $\tau_{o}$ (dashed grey line) and only start deviating at large values $\tau_{o}\approx2000$ ps.
On the other hand, when $\tau_{o}$ is kept constant at 200 ps and we change $\tau_{i}$, the values obtained from the fits seem to be determined by the value of the spin relaxation time in the outer regions.
From our calculations, this result holds for $D_{o}\leq D_{i}$, which falls in our experimental range.
%For $D_{o} > D_{i}$ we have a different trend, with the spin relaxation time being mostly limited by $\tau_{i}$.

The results for the spin relaxation time in our model can be explained in a qualitative manner.
Due to its diffusive motion, a part of the injected spins spend some time in the supported region before crossing the suspended region reaching the detector electrode (right boundary).
For a sufficiently low diffusion coefficient in the supported part, the time spent there is enough to relax most of the spins in the case of a short spin lifetime.
Considering $D_{s}\approx$0.1 m$^{2}$/s from our experimental results, the average time the spins take to diffuse through the central suspended region is about $\tau_{d}=\frac{L^{2}}{2D_{s}}\approx30$ ps for L=2.5 $\mu$m.
Therefore we should see a decrease in the effective spin relaxation time $\tau_{fit}$ in case the average time the spins take to diffuse through the central region is in the order of the spin relaxation time of this region ($\tau_{i}$), since most of the spins will relax before reaching the detector electrode.
This consequence can be seen in the right graph of Figure 3c.
If $\tau_{i}$ is longer than $\tau_{d}$, the main limiting factor in the effective relaxation time is $\tau_{o}$.

Now we will compare our theoretical results with the experimentally obtained values.
%From our simulation results we can now explain the values that we observe experimentally.
As said before, we have a good representation of D$_{s}$ for the central suspended region from the values extracted by the experimental Hanle curves.
The obtained values for the spin relaxation time, on the other hand, are apparently limited by the ``dirty'' outer regions.
By performing Hanle precession measurements in fully LOR supported spin-valves, we obtain values of $\tau_{s}\approx 150$ ps and $D_{s} \approx 0.02$ m$^{2}$/s, which support our conclusions.
This means that the spin relaxation time extracted from our Hanle precession measurements represents a lower bound on the spin relaxation time of the high quality central region.
%This effect can be seen as an ``impedance mismatch''\cite{PRB-Mihai2009,PRB-Wees2000} problem, in which the overall performance of the device is strongly affected by interconnects with different spintronic properties.
As far as we are aware this effect of inhomogeneity in a spin-valve was never reported and it is of importance to experiments in any inhomogeneous system and not exclusively for graphene.
For systematic studies in non-local spin valves as a function of the local properties of the central region (e.g. mobility, conductivity or carrier density) one must consider this inhomogeneity effects in the analysis, otherwise it may lead to erroneous interpretation of the experimental data.

Although we are not able to determine the actual spin relaxation time in the high mobility region, the experimental conditions allow us to take a very important conclusion with regard to the spin relaxation mechanism.
%We have shown that $\tau_{s}$ in the high mobility suspended graphene cannot be smaller than in the lower mobility supported regions.
If we consider the D'Yakonov-Perel as the main mechanism for spin relaxation in graphene, we can assume that $\tau_{s}$ relates with the momentum relaxation time $\tau_{p}$ according to \cite{SJETP-Dyakonov1971} Equation 1.
Since the charge diffusion coefficient relates to $\tau_{p}$ by: $D_{c}=\frac{v_{F}^2}{2}\tau_{p}$, we can extract for our high mobility samples $\tau_{p}\approx$0.2 ps.
We can then obtain an upper bound for the average spin-orbit coupling $\Delta_{SO} \leq$50 $\mu$eV using Equation 1.
It is important to notice that the value for $\Delta_{SO}$ that we estimate is an upper bound which is calculated from the experimentally determined lower bound on the spin relaxation time in the suspended high quality graphene.
Applying the same procedure for typical SiO$_{2}$ supported graphene spin valves: $\tau_{p}$=0.04 ps and $\tau_{s}$=200 ps we obtain an average SO coupling of $\Delta_{SO}$= 110 $\mu$eV.
If this value was intrinsic for graphene we would obtain a strong reduction in the spin relaxation time in our high quality graphene devices down to $\tau_{s}$=50 ps.
This value for $\tau_{s}$ is comparable to the time the spins take to diffuse through the central suspended region ($\tau_{d}$) and smaller than the spin relaxation time for the non-suspended regions.
This means that if the value for SO coupling obtained in the SiO$_{2}$ devices was an intrinsic value for graphene we should be able to observe a very low $\tau_{s}$, which is not the case.
From this result we can conclude that the nature of the SO fields in the graphene devices observed so-far is not due to graphene's properties, but due to extrinsic effects.
The value obtained for $\Delta_{SO}$ in our suspended graphene devices is about a factor of two lower than what is usually obtained in the low mobility SiO$_{2}$ supported devices, although it is still five times higher than the theoretical limit of $\Delta_{SO} \approx$10 $\mu$eV \cite{PRL-Brataas2009,PRB-MacDonald2006,PRB-Fabian2009}.
Our results are in agreement with recent theoretical predictions that $\Delta_{SO}$ scales with the presence of adatoms\cite{PRL-Castroneto2009,NJP-Wu2012}, so higher spin lifetimes in high quality graphene flakes should be expected.

%\section{Conclusions}
In conclusion we performed spin transport in suspended high mobility monolayer graphene devices contacted by ferromagnetic leads.
We showed that electronic mobilities above 100,000 cm$^{2}$/Vs for the suspended region can be achieved in our devices via current annealing.
Our measurements showed an increase up to one order of magnitude in the spin diffusion coefficient when compared to SiO$_{2}$ supported devices and very large spin relaxation lengths ($\lambda$ = 4.7 $\mu$m) were obtained.
We did not observe a change in the measured spin relaxation time when comparing our results to the traditional SiO$_{2}$ supported devices.
This effect is explained by considering a simple model that takes the inhomogeneity in our devices into account, considering not only the suspended regions of the devices, but also the supported part.
We observe that the spin transport measurements of our high mobility graphene device strongly depends on how spins interact in the lower mobility graphene parts directly connected to it.
In a similar way this inhomogeneity effect is expected to strongly affect the performance of a typical graphene nanodevice when connected to other graphene nanodevices or interconnects with different spintronic properties (due to possible edge roughness or other kind of spin scattering in the system).
For future works, both in graphene and non-graphene based devices, one must take this effect into account in order to make concise conclusions.
We were also able to give a higher bound for the spin-orbit coupling in our devices of 50 $\mu$eV, a factor of 2 lower than of those encountered in SiO$_{2}$ supported devices.

%% classes.  Rather than use \section, an appropriate macro is
%% provided that will always work.
%%%%%%%%%%%%%%%%%%%%%%%%%%%%%%%%%%%%%%%%%%%%%%%%%%%%%%%%%%%%%%%%%%%%%
\acknowledgement

We would like to acknowledge J. G. Holstein, B. Wolfs and H. M. de Roosz for their technical support and M. W. Wu for discussions. This work was realized using NanoLab.NL (NanoNed) facilities and is part of the research program of the Foundation for Fundamental Research on Matter (FOM), which is part of the Netherlands Organization for Scientific Research (NWO). This work was partially supported by the Zernike Institute for Advanced Materials.

%% \suppinfo macro.
%%%%%%%%%%%%%%%%%%%%%%%%%%%%%%%%%%%%%%%%%%%%%%%%%%%%%%%%%%%%%%%%%%%%%

\suppinfo
\section{Device Preparation}
The devices are prepared in a similar fashion as the ones described by Tombros et al. \cite{JAP-Niko2011}, although a few changes were made in order to avoid degradation of our high resistive contact barriers.
First a 1 $\mu$m thick lift-off resist (LOR) film is spin-coated on a Si/SiO$_{2}$ (500 nm) substrate and the graphene flakes are exfoliated on top.
Single layer flakes are then selected by optical contrast using a green filter \cite{APL-Geim2007,APL-Falko2007}.
For the electron beam lithography (EBL) process, to improve the undercut, we use a double layer Polymethyl methacrylate (PMMA) 50/410K resists dissolved in chlorobenzene and o-Xylene respectively.
These solvents are used to prevent the removal of the LOR film during spin-coating.
The contacts are then patterned and developed in n-Xylene (20$^{o}$ C).

Using an electron beam evaporator with a base pressure lower than $8\times10^{-7}$ Torr, we deposit 0.4 nm of Aluminium followed by \textit{in-situ} oxidation by pure Oxygen gas at a pressure higher than $1\times10^{-2}$ Torr for 15 minutes and the chamber is pumped down to the initial base pressure.
This process is performed twice in order to get contact resistances higher than 10 k$\Omega$.
After the high resistance barriers are deposited the chamber is pumped down to the initial base pressure and 60 nm of Cobalt is evaporated.
For some of the studied samples in this work the electrodes were capped by 3 nm of Al$_{2}$O$_{3}$ to prevent Co oxidation.
The lift-off is done in hot (75$^{o}$ C) n-Xylene.

To suspend the graphene flakes a second EBL step is performed with an area dose of 510 $\mu$C/cm$^{2}$ and developed in 1-methyl 2-propanol.
It was found that if the sample is immersed in Ethyl-lactat as described by Tombros et al. \cite{JAP-Niko2011} the AlOx barriers degrade, causing a very large increase in the contact resistance and loss of the spin-signal.
After this final process the sample is bonded and loaded in a cryostat which is pumped down to a base pressure lower than $1\times10^{-6}$ Torr.

%%%%%%%%%%%%%%%%%%%%%%%%%%%%%%%%%%%%%%%%%%%%%%%%%%%%%%%%%%%%%%%%%%%%%%%%%%%%%%%%%%%%%%
\section{Current annealing}
After the sample is loaded in the cryostat and characterized, we perform a current annealing step to remove the impurities in the graphene flake and obtain a high mobility device.
The whole procedure is carried at 4.2 K.

To avoid degradation of the electrodes used for spin injection/detection, we apply the large DC bias for the current annealing in the two outer electrodes as depicted in Figure 4. The contact resistance of the inner contacts were measured before and after the current annealing step and showed no noticeable change.

\begin{figure}
	\centering
		\includegraphics[width=0.5\textwidth]{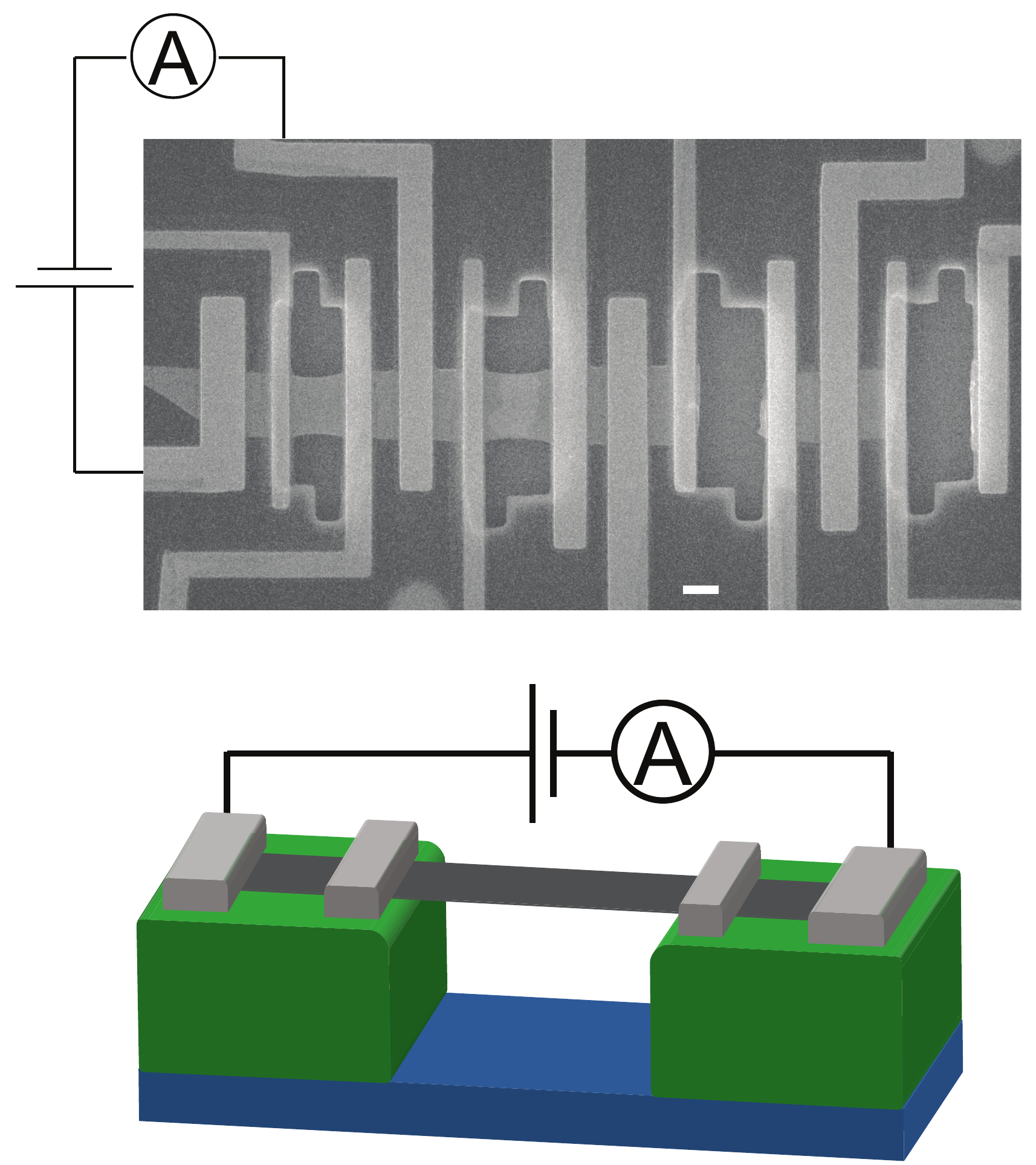}
	\caption{Top: A SEM picture of a typical device showing the schematics for the current annealing setup. On the left of the picture it can be seen two regions of suspended graphene flake, the outermost left successfully cleaned. On the right it can be seen two broken regions due to failing in the current annealing procedure. The scale bar measures 1 $\mu$m. Bottom: A cartoon illustrating the current annealing setup for comparison.}
	\label{fig:si-ca}
\end{figure}

We use a DC current bias-voltage compliance procedure to limit the power in our devices and avoid them to burn.
The current is ramped up slowly ($approx$ 1 $\mu$A/s) until a determined value and then rapidly ramped down, at a rate 4 times faster than the ramping up.
After each sweep in current we check the gate-voltage dependence on the sheet resistance to keep track of the device's mobility.
For our samples, this current annealing procedure had a success rate of about 33\% (4 out of 12 regions showed high mobility), comparable with our previous results \cite{JAP-Niko2011}.
One of the regions was current annealed twice.
The first procedure resulted in a mobility of $\mu \approx$ 10$^{5}$ cm$^{2}$/Vs, and after the second current annealing the mobility was improved to $\mu \approx$ 3 $\times$ 10$^{5}$ cm$^{2}$/Vs.
Despite performing spin-transport measurements in this sample, we also obtained spin signals in another sample with similar properties.

%%%%%%%%%%%%%%%%%%%%%%%%%%%%%%%%%%%%%%%%%%%%%%%%%%%%%%%%%%%%%%%%%%%%%%%%%%%%%%%%%%%%%%
\section{Details on the simulation}
To represent our sample we extended the 1D model by Popinciuc et al. \cite{PRB-Mihai2009} to include three different regions: two semi-infinite outer parts of width W sandwiching one inner part of length L and width W.
In this model we solve the stationary Bloch equations for the three components of the spin accumulation $\vec{\mu}_{s}(x)$ in the presence of a magnetic field $\vec{B}$:

\begin{equation}
\label{eq:bloch}
D_{s} \nabla^{2} \vec{\mu}_{s} - \frac{\vec{\mu}_{s}}{\tau_{s}} + \gamma \vec{B} \times \vec{\mu_{s}} = 0
\end{equation}

\noindent where $D_{s}$ is the spin diffusion constant, $\tau_{s}$ is the spin relaxation time and $\gamma=g \mu_{b} \hbar^{-1}$ is the gyromagnetic ratio.
All the parameters of the equation above can be set separately for each of the three regions, but for simplicity we make the two outer regions identical.
For the boundary conditions we take:
\begin{enumerate}
	\item $\mu_{s}(x=\pm \infty)=0$
	\item $\mu_{s}(x=0_{+})=\mu_{s}(x=0_{-})$
	\item $\mu_{s}(x=L_{+})=\mu_{s}(x=L_{-})$
	\item $\beta = \frac{\sigma_{o}W}{2e} \frac{d \mu_{s}(x=0_{-})}{dx} - \frac{\sigma_{i}W}{2e} \frac{d \mu_{s}(x=0_{+})}{dx} + \frac{\mu_{s}(x=0)}{2eR_{c1}}$
	\item $0 = \frac{\sigma_{i}W}{2e} \frac{d \mu_{s}(x=L_{-})}{dx} - \frac{\sigma_{o}W}{2e} \frac{d \mu_{s}(x=L_{+})}{dx} + \frac{\mu_{s}(x=L)}{2eR_{c2}}$
\end{enumerate}
\noindent where $\beta = \frac{PI}{2}$ for the x component of $\mu_{s}$ and $\beta = 0$ for the y and z components, with $P$ being the spin polarization of the charge current $I$ injected in the left boundary.
The contact resistance of the contact at the left (right) boundary is represented by $R_{c1(c2)}$, and the conductivity of the inner (outer) regions by $\sigma_{i(o)}$.
%that the spin accumulation goes to zero at $x=\pm \infty$, $\vec{\mu}_{s}$ and its derivatives are continuous at the boundaries except for the derivative of the $x$ component in the left boundary, $\mu_{s}^{x}(x=0)$, that sets the initial spin accumulation created by a spin injector.
%The contact induced spin relaxation is included as described in Ref. \cite{PRB-Mihai2009}, although contact effects were found to be negligible for our results when we consider values obtained experimentally in our samples.
The contact induced spin relaxation is represented by the last term of items 4 and 5 \cite{PRB-Mihai2009}, although contact effects were found to be negligible for our results when we consider values obtained experimentally in our samples.

We have to study the effect of four different parameters: the spin diffusion constants $D_{i}$ and $D_{o}$, and the spin relaxation times $\tau_{i}$ and $\tau_{o}$, where the subscripts "i" and "o" refer to the inner and outer regions respectively.
In order to be able to observe the effect of each one of the parameters separately we calculated several Hanle precession curves keeping three of them constant and vary the remaining one.
The simulated precession curves for a few sets of parameters are depicted in Figure 5.
These curves were then fitted using the solution for the Bloch equations in a homogeneous system, like we fit our experimental results.
From these fits we obtain an "effective" spin diffusion constant $D_{fit}$ and relaxation time $\tau_{fit}$ as shown in the main manuscript.
We also tried to fit our experimental data with the curves we get from our model, but it lead to similar results to the ones obtained using the solution for a homogeneous system.
It is worth noting that when we compare the curves in Figure 5a, we observe that in the case of changing $D_{o}$ the obtained precession curves change in magnitude but not in shape.
This means that we do not observe any change in the values obtained for $D_{fit}$ or $\tau_{fit}$.
On the other hand, analyzing Figure 5b we see that changes in $D_{i}$ does not only change the magnitude of the spin signal but also the shape of the curve, which results in changes in $D_{fit}$ with changes in $D_{i}$.
A similar effect is observed in the analysis of Figure 5c and Figure 5d.

\begin{figure}
	\centering
		\includegraphics[width=1.0\textwidth]{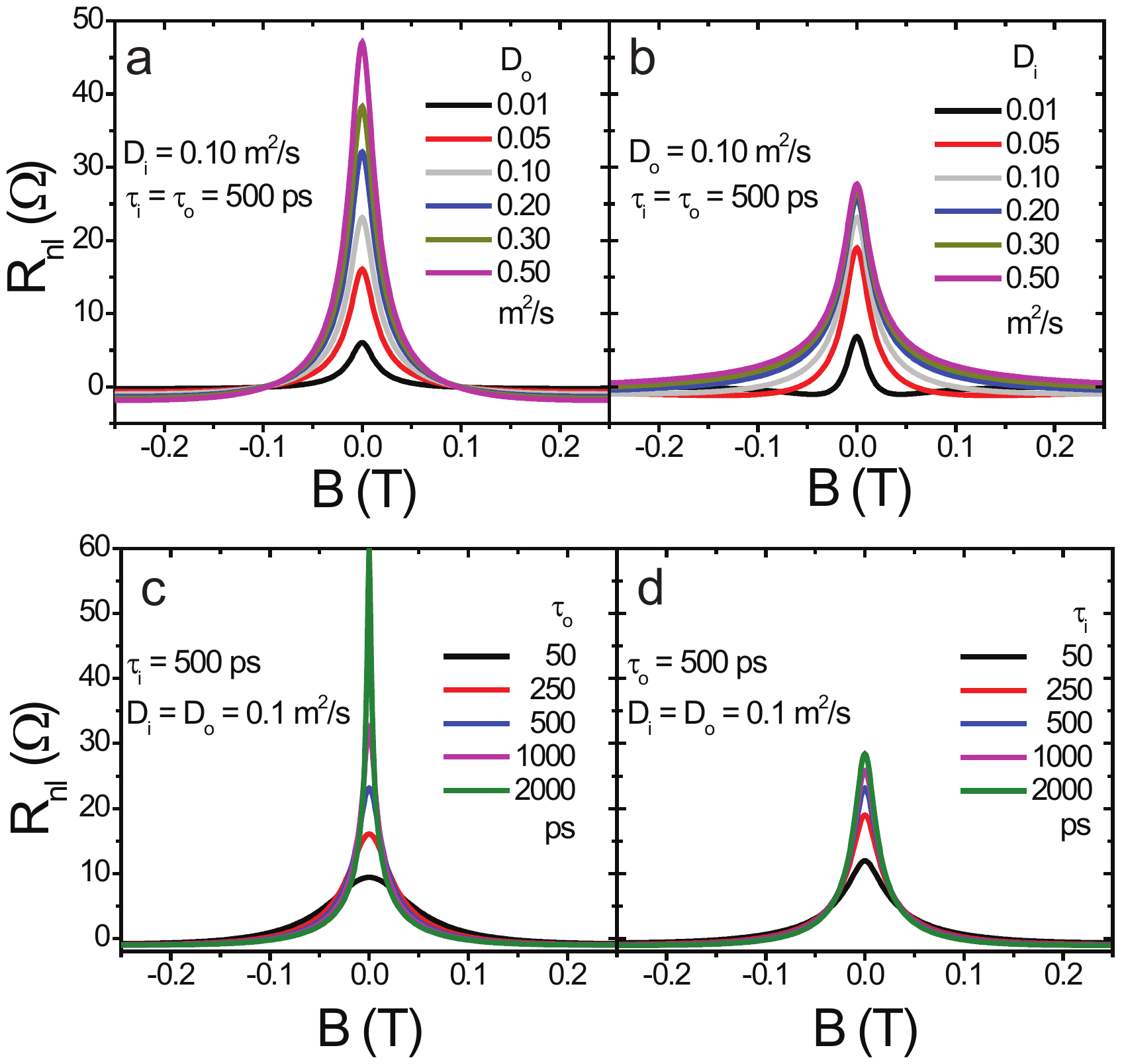}
	\caption{The calculated Hanle precession curves for different values of (a)$D_{o}$, (b)$D_{i}$, (c)$\tau_{o}$ and (d)$\tau_{i}$, while keeping the other parameters fixed.}
	\label{fig:si-change}
\end{figure}

%%%%%%%%%%%%%%%%%%%%%%%%%%%%%%%%%%%%%%%%%%%%%%%%%%%%%%%%%%%%%%%%%%%%%%%%%%%%%%%%%%%%%%
\section{The effect of the sheet resistance on the Hanle precession}

The same way as we can change the values for the spin diffusion coefficients and relaxation times for the inner and outer parts ($D_{o}$, $D_{i}$, $\tau_{o}$ and $\tau_{i}$) we can also change the values for the square resistances of the inner and outer regions, $R_{i}$ and $R_{o}$ respectively.
By applying the same procedure of generating a Hanle precession curve and fitting it with the homogeneous model we can extract the effective spin relaxation time $\tau_{fit}$ and the effective spin diffusion coefficient $D_{fit}$.
The results for different combinations of $R_{i}$ and $R_{o}$ are presented in Figure 6 below.

\begin{figure}
	\centering
		\includegraphics[width=0.5\textwidth]{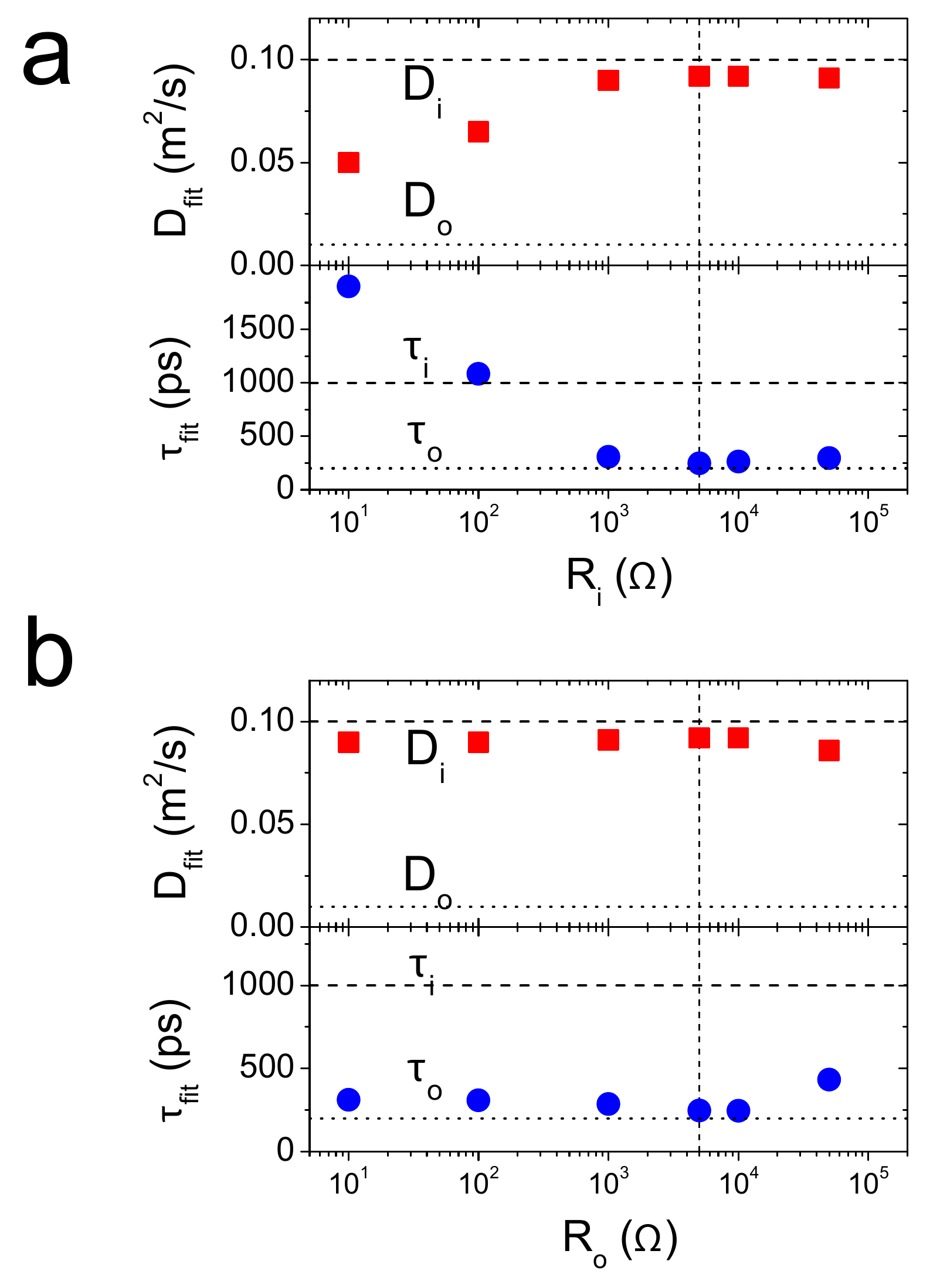}
	\caption{(a) The spin signal obtained for different values of the inner (outer) sheet resistance while keeping the outer (inner) constant at 5 k$\Omega$. The results obtained for $D_{fit}$ and $\tau_{fit}$ for changing (b)$R_{i}$ and (c)$R_{o}$. The outer or inner sheet resistance was kept at 5 k$\Omega$ while the other was changed. The values for the diffusion coefficients and spin relaxation times are: $D_{o}$=0.01 m$^{2}$/s, $D_{i}$0.1 m$^{2}$/s, $\tau_{o}$= 200 ps and $\tau_{i}$= 1 ps.}
	\label{fig:SI-Rsq}
\end{figure}

As it can be seen in Figure 6 (b) and (c), the conclusions obtained in our main text that $D_{fit}$ is determined mainly by the inner region and $\tau_{fit}$ by the outer region remain unchanged when we consider resistances in the range of those we encounter experimentally (1 to 5 k$\Omega$).

%\bibliography{ssv-references}

\begin{mcitethebibliography}{36}
\providecommand*\natexlab[1]{#1}
\providecommand*\mciteSetBstSublistMode[1]{}
\providecommand*\mciteSetBstMaxWidthForm[2]{}
\providecommand*\mciteBstWouldAddEndPuncttrue
  {\def\EndOfBibitem{\unskip.}}
\providecommand*\mciteBstWouldAddEndPunctfalse
  {\let\EndOfBibitem\relax}
\providecommand*\mciteSetBstMidEndSepPunct[3]{}
\providecommand*\mciteSetBstSublistLabelBeginEnd[3]{}
\providecommand*\EndOfBibitem{}
\mciteSetBstSublistMode{f}
\mciteSetBstMaxWidthForm{subitem}{(\alph{mcitesubitemcount})}
\mciteSetBstSublistLabelBeginEnd
  {\mcitemaxwidthsubitemform\space}
  {\relax}
  {\relax}

\bibitem[Huertas-Hernando et~al.(2007)Huertas-Hernando, Guinea, and
  Brataas]{EPJ-Brataas2007}
Huertas-Hernando,~D.; Guinea,~F.; Brataas,~A. \emph{The European Physical
  Journal - Special Topics} \textbf{2007}, \emph{148}, 177--181,
  10.1140/epjst/e2007-00238-0\relax
\mciteBstWouldAddEndPuncttrue
\mciteSetBstMidEndSepPunct{\mcitedefaultmidpunct}
{\mcitedefaultendpunct}{\mcitedefaultseppunct}\relax
\EndOfBibitem
\bibitem[Huertas-Hernando et~al.(2009)Huertas-Hernando, Guinea, and
  Brataas]{PRL-Brataas2009}
Huertas-Hernando,~D.; Guinea,~F.; Brataas,~A. \emph{Phys. Rev. Lett.}
  \textbf{2009}, \emph{103}, 146801\relax
\mciteBstWouldAddEndPuncttrue
\mciteSetBstMidEndSepPunct{\mcitedefaultmidpunct}
{\mcitedefaultendpunct}{\mcitedefaultseppunct}\relax
\EndOfBibitem
\bibitem[J\'{o}zsa et~al.(2009)J\'{o}zsa, Maassen, Popinciuc, Zomer, Veligura,
  Jonkman, and van Wees]{PRB-Csaba2009}
J\'{o}zsa,~C.; Maassen,~T.; Popinciuc,~M.; Zomer,~P.~J.; Veligura,~A.;
  Jonkman,~H.~T.; van Wees,~B.~J. \emph{Physical Review B (Condensed Matter and
  Materials Physics)} \textbf{2009}, \emph{80}, 241403\relax
\mciteBstWouldAddEndPuncttrue
\mciteSetBstMidEndSepPunct{\mcitedefaultmidpunct}
{\mcitedefaultendpunct}{\mcitedefaultseppunct}\relax
\EndOfBibitem
\bibitem[Han et~al.(2010)Han, Pi, McCreary, Li, Wong, Swartz, and
  Kawakami]{PRL-Kawakami2010}
Han,~W.; Pi,~K.; McCreary,~K.~M.; Li,~Y.; Wong,~J. J.~I.; Swartz,~A.~G.;
  Kawakami,~R.~K. \emph{Phys. Rev. Lett.} \textbf{2010}, \emph{105},
  167202\relax
\mciteBstWouldAddEndPuncttrue
\mciteSetBstMidEndSepPunct{\mcitedefaultmidpunct}
{\mcitedefaultendpunct}{\mcitedefaultseppunct}\relax
\EndOfBibitem
\bibitem[Avsar et~al.(2011)Avsar, Yang, Bae, Balakrishnan, Volmer, Jaiswal, Yi,
  Ali, Gu¨ntherodt, Hong, Beschoten, and O¨zyilmaz]{Nanoletters-Barbaros2011}
Avsar,~A.; Yang,~T.-Y.; Bae,~S.; Balakrishnan,~J.; Volmer,~F.; Jaiswal,~M.;
  Yi,~Z.; Ali,~S.~R.; Gu¨ntherodt,~G.; Hong,~B.~H.; Beschoten,~B.;
  O¨zyilmaz,~B. \emph{Nano Letters} \textbf{2011}, \emph{11}, 2363--2368\relax
\mciteBstWouldAddEndPuncttrue
\mciteSetBstMidEndSepPunct{\mcitedefaultmidpunct}
{\mcitedefaultendpunct}{\mcitedefaultseppunct}\relax
\EndOfBibitem
\bibitem[Tombros et~al.(2007)Tombros, Jozsa, Popinciuc, Jonkman, and van
  Wees]{Nature-Niko2007}
Tombros,~N.; Jozsa,~C.; Popinciuc,~M.; Jonkman,~H.~T.; van Wees,~B.~J.
  \emph{Nature} \textbf{2007}, \emph{448}, 571--574\relax
\mciteBstWouldAddEndPuncttrue
\mciteSetBstMidEndSepPunct{\mcitedefaultmidpunct}
{\mcitedefaultendpunct}{\mcitedefaultseppunct}\relax
\EndOfBibitem
\bibitem[Popinciuc et~al.(2009)Popinciuc, J\'ozsa, Zomer, Tombros, Veligura,
  Jonkman, and van Wees]{PRB-Mihai2009}
Popinciuc,~M.; J\'ozsa,~C.; Zomer,~P.~J.; Tombros,~N.; Veligura,~A.;
  Jonkman,~H.~T.; van Wees,~B.~J. \emph{Phys. Rev. B} \textbf{2009}, \emph{80},
  214427\relax
\mciteBstWouldAddEndPuncttrue
\mciteSetBstMidEndSepPunct{\mcitedefaultmidpunct}
{\mcitedefaultendpunct}{\mcitedefaultseppunct}\relax
\EndOfBibitem
\bibitem[Han and Kawakami(2011)Han, and Kawakami]{PRL-Kawakami2011}
Han,~W.; Kawakami,~R.~K. \emph{Phys. Rev. Lett.} \textbf{2011}, \emph{107},
  047207\relax
\mciteBstWouldAddEndPuncttrue
\mciteSetBstMidEndSepPunct{\mcitedefaultmidpunct}
{\mcitedefaultendpunct}{\mcitedefaultseppunct}\relax
\EndOfBibitem
\bibitem[Maassen et~al.(2011)Maassen, Dejene, Guimar\~aes, J\'ozsa, and van
  Wees]{PRB-Thomas2011}
Maassen,~T.; Dejene,~F.~K.; Guimar\~aes,~M. H.~D.; J\'ozsa,~C.; van Wees,~B.~J.
  \emph{Phys. Rev. B} \textbf{2011}, \emph{83}, 115410\relax
\mciteBstWouldAddEndPuncttrue
\mciteSetBstMidEndSepPunct{\mcitedefaultmidpunct}
{\mcitedefaultendpunct}{\mcitedefaultseppunct}\relax
\EndOfBibitem
\bibitem[Castro~Neto and Guinea(2009)Castro~Neto, and
  Guinea]{PRL-Castroneto2009}
Castro~Neto,~A.~H.; Guinea,~F. \emph{Phys. Rev. Lett.} \textbf{2009},
  \emph{103}, 026804\relax
\mciteBstWouldAddEndPuncttrue
\mciteSetBstMidEndSepPunct{\mcitedefaultmidpunct}
{\mcitedefaultendpunct}{\mcitedefaultseppunct}\relax
\EndOfBibitem
\bibitem[Ochoa et~al.(2012)Ochoa, Castro~Neto, and Guinea]{PRL-Guinea2012}
Ochoa,~H.; Castro~Neto,~A.~H.; Guinea,~F. \emph{Phys. Rev. Lett.}
  \textbf{2012}, \emph{108}, 206808\relax
\mciteBstWouldAddEndPuncttrue
\mciteSetBstMidEndSepPunct{\mcitedefaultmidpunct}
{\mcitedefaultendpunct}{\mcitedefaultseppunct}\relax
\EndOfBibitem
\bibitem[Zhang and Wu(2012)Zhang, and Wu]{NJP-Wu2012}
Zhang,~P.; Wu,~M.~W. \emph{New Journal of Physics} \textbf{2012}, \emph{14},
  033015\relax
\mciteBstWouldAddEndPuncttrue
\mciteSetBstMidEndSepPunct{\mcitedefaultmidpunct}
{\mcitedefaultendpunct}{\mcitedefaultseppunct}\relax
\EndOfBibitem
\bibitem[{D'Yakonov} and {Perel'}(1971){D'Yakonov}, and
  {Perel'}]{SJETP-Dyakonov1971}
{D'Yakonov},~M.~I.; {Perel'},~V.~I. \emph{Soviet Journal of Experimental and
  Theoretical Physics} \textbf{1971}, \emph{33}, 1053\relax
\mciteBstWouldAddEndPuncttrue
\mciteSetBstMidEndSepPunct{\mcitedefaultmidpunct}
{\mcitedefaultendpunct}{\mcitedefaultseppunct}\relax
\EndOfBibitem
\bibitem[Maassen et~al.(0)Maassen, van~den Berg, IJbema, Fromm, Seyller,
  Yakimova, and van Wees]{Nanolett-Thomas2012}
Maassen,~T.; van~den Berg,~J.~J.; IJbema,~N.; Fromm,~F.; Seyller,~T.;
  Yakimova,~R.; van Wees,~B.~J. \emph{Nano Letters} \textbf{0}, \emph{0},
  null\relax
\mciteBstWouldAddEndPuncttrue
\mciteSetBstMidEndSepPunct{\mcitedefaultmidpunct}
{\mcitedefaultendpunct}{\mcitedefaultseppunct}\relax
\EndOfBibitem
\bibitem[Zhang et~al.(2009)Zhang, Brar, Girit, Zettl, and
  Crommie]{Natphys-Crommie2009}
Zhang,~Y.; Brar,~V.~W.; Girit,~C.; Zettl,~A.; Crommie,~M.~F. \emph{Nat Phys}
  \textbf{2009}, \emph{5}, 722--726\relax
\mciteBstWouldAddEndPuncttrue
\mciteSetBstMidEndSepPunct{\mcitedefaultmidpunct}
{\mcitedefaultendpunct}{\mcitedefaultseppunct}\relax
\EndOfBibitem
\bibitem[Bolotin et~al.(2008)Bolotin, Sikes, Jiang, Klima, Fudenberg, Hone,
  Kim, and Stormer]{SSC-Kim2008}
Bolotin,~K.; Sikes,~K.; Jiang,~Z.; Klima,~M.; Fudenberg,~G.; Hone,~J.; Kim,~P.;
  Stormer,~H. \emph{Solid State Communications} \textbf{2008}, \emph{146}, 351
  -- 355\relax
\mciteBstWouldAddEndPuncttrue
\mciteSetBstMidEndSepPunct{\mcitedefaultmidpunct}
{\mcitedefaultendpunct}{\mcitedefaultseppunct}\relax
\EndOfBibitem
\bibitem[Du et~al.(2008)Du, Skachko, Barker, and Andrei]{Natnano-Andrei2008}
Du,~X.; Skachko,~I.; Barker,~A.; Andrei,~E.~Y. \emph{Nat Nano} \textbf{2008},
  \emph{3}, 491--495\relax
\mciteBstWouldAddEndPuncttrue
\mciteSetBstMidEndSepPunct{\mcitedefaultmidpunct}
{\mcitedefaultendpunct}{\mcitedefaultseppunct}\relax
\EndOfBibitem
\bibitem[Garcia-Sanchez et~al.(2008)Garcia-Sanchez, van~der Zande, Paulo,
  Lassagne, McEuen, and Bachtold]{Nanolett-Mceuen2008}
Garcia-Sanchez,~D.; van~der Zande,~A.~M.; Paulo,~A.~S.; Lassagne,~B.;
  McEuen,~P.~L.; Bachtold,~A. \emph{Nano Letters} \textbf{2008}, \emph{8},
  1399--1403, PMID: 18402478\relax
\mciteBstWouldAddEndPuncttrue
\mciteSetBstMidEndSepPunct{\mcitedefaultmidpunct}
{\mcitedefaultendpunct}{\mcitedefaultseppunct}\relax
\EndOfBibitem
\bibitem[Guinea et~al.(2010)Guinea, Katsnelson, and Geim]{Natphys-Geim2010}
Guinea,~F.; Katsnelson,~M.~I.; Geim,~A.~K. \emph{Nat Phys} \textbf{2010},
  \emph{6}, 30--33\relax
\mciteBstWouldAddEndPuncttrue
\mciteSetBstMidEndSepPunct{\mcitedefaultmidpunct}
{\mcitedefaultendpunct}{\mcitedefaultseppunct}\relax
\EndOfBibitem
\bibitem[Levy et~al.(2010)Levy, Burke, Meaker, Panlasigui, Zettl, Guinea, Neto,
  and Crommie]{Science-Crommie2010}
Levy,~N.; Burke,~S.~A.; Meaker,~K.~L.; Panlasigui,~M.; Zettl,~A.; Guinea,~F.;
  Neto,~A. H.~C.; Crommie,~M.~F. \emph{Science} \textbf{2010}, \emph{329},
  544--547\relax
\mciteBstWouldAddEndPuncttrue
\mciteSetBstMidEndSepPunct{\mcitedefaultmidpunct}
{\mcitedefaultendpunct}{\mcitedefaultseppunct}\relax
\EndOfBibitem
\bibitem[Zolfagharkhani et~al.(2008)Zolfagharkhani, Gaidarzhy, Degiovanni,
  Kettemann, Fulde, and Mohanty]{NatNano-Mohanty2008}
Zolfagharkhani,~G.; Gaidarzhy,~A.; Degiovanni,~P.; Kettemann,~S.; Fulde,~P.;
  Mohanty,~P. \emph{Nat Nano} \textbf{2008}, \emph{3}, 720--723\relax
\mciteBstWouldAddEndPuncttrue
\mciteSetBstMidEndSepPunct{\mcitedefaultmidpunct}
{\mcitedefaultendpunct}{\mcitedefaultseppunct}\relax
\EndOfBibitem
\bibitem[Haney and Stiles(2010)Haney, and Stiles]{PRL-Stiles2010}
Haney,~P.~M.; Stiles,~M.~D. \emph{Phys. Rev. Lett.} \textbf{2010}, \emph{105},
  126602\relax
\mciteBstWouldAddEndPuncttrue
\mciteSetBstMidEndSepPunct{\mcitedefaultmidpunct}
{\mcitedefaultendpunct}{\mcitedefaultseppunct}\relax
\EndOfBibitem
\bibitem[Kovalev et~al.(2011)Kovalev, Hayden, Bauer, and
  Tserkovnyak]{PRL-Kovalev2011}
Kovalev,~A.~A.; Hayden,~L.~X.; Bauer,~G. E.~W.; Tserkovnyak,~Y. \emph{Phys.
  Rev. Lett.} \textbf{2011}, \emph{106}, 147203\relax
\mciteBstWouldAddEndPuncttrue
\mciteSetBstMidEndSepPunct{\mcitedefaultmidpunct}
{\mcitedefaultendpunct}{\mcitedefaultseppunct}\relax
\EndOfBibitem
\bibitem[Not()]{Note-1}
Up to now all previous results for graphene spin valves on SiO$_{2}$ or for our
  suspended devices without cleaning show a constant behaviour of
  $\frac{\tau_{s}}{\tau_{p}}$ as a function of the Fermi energy. This goes
  against the theories for Elliott-Yafet mechanism for spin relaxation in
  graphene as showed in Ref. 11. This way we assume the most recent theory that
  best describe all the measurements of spin transport in graphene available
  so-far (Ref. 12).\relax
\mciteBstWouldAddEndPunctfalse
\mciteSetBstMidEndSepPunct{\mcitedefaultmidpunct}
{}{\mcitedefaultseppunct}\relax
\EndOfBibitem
\bibitem[Min et~al.(2006)Min, Hill, Sinitsyn, Sahu, Kleinman, and
  MacDonald]{PRB-MacDonald2006}
Min,~H.; Hill,~J.~E.; Sinitsyn,~N.~A.; Sahu,~B.~R.; Kleinman,~L.;
  MacDonald,~A.~H. \emph{Phys. Rev. B} \textbf{2006}, \emph{74}, 165310\relax
\mciteBstWouldAddEndPuncttrue
\mciteSetBstMidEndSepPunct{\mcitedefaultmidpunct}
{\mcitedefaultendpunct}{\mcitedefaultseppunct}\relax
\EndOfBibitem
\bibitem[Gmitra et~al.(2009)Gmitra, Konschuh, Ertler, Ambrosch-Draxl, and
  Fabian]{PRB-Fabian2009}
Gmitra,~M.; Konschuh,~S.; Ertler,~C.; Ambrosch-Draxl,~C.; Fabian,~J.
  \emph{Phys. Rev. B} \textbf{2009}, \emph{80}, 235431\relax
\mciteBstWouldAddEndPuncttrue
\mciteSetBstMidEndSepPunct{\mcitedefaultmidpunct}
{\mcitedefaultendpunct}{\mcitedefaultseppunct}\relax
\EndOfBibitem
\bibitem[Fabian et~al.(2007)Fabian, Matos-Abiague, Ertler, Stano, and
  Zutic]{APS-Fabian2007}
Fabian,~J.; Matos-Abiague,~A.; Ertler,~C.; Stano,~P.; Zutic,~I. \emph{Acta
  Physica Slovaca} \textbf{2007}, \emph{57}, 565--907\relax
\mciteBstWouldAddEndPuncttrue
\mciteSetBstMidEndSepPunct{\mcitedefaultmidpunct}
{\mcitedefaultendpunct}{\mcitedefaultseppunct}\relax
\EndOfBibitem
\bibitem[Tombros et~al.(2011)Tombros, Veligura, Junesch, van~den Berg, Zomer,
  Wojtaszek, Vera-Marun, Jonkman, and van Wees]{JAP-Niko2011}
Tombros,~N.; Veligura,~A.; Junesch,~J.; van~den Berg,~J.~J.; Zomer,~P.~J.;
  Wojtaszek,~M.; Vera-Marun,~I.~J.; Jonkman,~H.~T.; van Wees,~B.~J.
  \emph{Journal of Applied Physics} \textbf{2011}, \emph{109}, 093702\relax
\mciteBstWouldAddEndPuncttrue
\mciteSetBstMidEndSepPunct{\mcitedefaultmidpunct}
{\mcitedefaultendpunct}{\mcitedefaultseppunct}\relax
\EndOfBibitem
\bibitem[Blake et~al.(2007)Blake, Hill, Neto, Novoselov, Jiang, Yang, Booth,
  and Geim]{APL-Geim2007}
Blake,~P.; Hill,~E.~W.; Neto,~A. H.~C.; Novoselov,~K.~S.; Jiang,~D.; Yang,~R.;
  Booth,~T.~J.; Geim,~A.~K. \emph{Applied Physics Letters} \textbf{2007},
  \emph{91}, 063124\relax
\mciteBstWouldAddEndPuncttrue
\mciteSetBstMidEndSepPunct{\mcitedefaultmidpunct}
{\mcitedefaultendpunct}{\mcitedefaultseppunct}\relax
\EndOfBibitem
\bibitem[Abergel et~al.(2007)Abergel, Russell, and Fal'ko]{APL-Falko2007}
Abergel,~D. S.~L.; Russell,~A.; Fal'ko,~V.~I. \emph{Applied Physics Letters}
  \textbf{2007}, \emph{91}, 063125\relax
\mciteBstWouldAddEndPuncttrue
\mciteSetBstMidEndSepPunct{\mcitedefaultmidpunct}
{\mcitedefaultendpunct}{\mcitedefaultseppunct}\relax
\EndOfBibitem
\bibitem[Tombros et~al.(2011)Tombros, Veligura, Junesch, Guimaraes, Vera-Marun,
  Jonkman, and van Wees]{NatPhys-Niko2011}
Tombros,~N.; Veligura,~A.; Junesch,~J.; Guimaraes,~M. H.~D.; Vera-Marun,~I.~J.;
  Jonkman,~H.~T.; van Wees,~B.~J. \emph{Nat Phys} \textbf{2011}, \emph{7},
  697--700\relax
\mciteBstWouldAddEndPuncttrue
\mciteSetBstMidEndSepPunct{\mcitedefaultmidpunct}
{\mcitedefaultendpunct}{\mcitedefaultseppunct}\relax
\EndOfBibitem
\bibitem[Moser et~al.(2007)Moser, Barreiro, and Bachtold]{APL-Bachtold2007}
Moser,~J.; Barreiro,~A.; Bachtold,~A. \emph{Applied Physics Letters}
  \textbf{2007}, \emph{91}, 163513\relax
\mciteBstWouldAddEndPuncttrue
\mciteSetBstMidEndSepPunct{\mcitedefaultmidpunct}
{\mcitedefaultendpunct}{\mcitedefaultseppunct}\relax
\EndOfBibitem
\bibitem[Schmidt et~al.(2000)Schmidt, Ferrand, Molenkamp, Filip, and van
  Wees]{PRB-Wees2000}
Schmidt,~G.; Ferrand,~D.; Molenkamp,~L.~W.; Filip,~A.~T.; van Wees,~B.~J.
  \emph{Phys. Rev. B} \textbf{2000}, \emph{62}, R4790--R4793\relax
\mciteBstWouldAddEndPuncttrue
\mciteSetBstMidEndSepPunct{\mcitedefaultmidpunct}
{\mcitedefaultendpunct}{\mcitedefaultseppunct}\relax
\EndOfBibitem
\bibitem[Hwang et~al.(2007)Hwang, Adam, and Das~Sarma]{PRL-Sarma2007}
Hwang,~E.~H.; Adam,~S.; Das~Sarma,~S. \emph{Phys. Rev. Lett.} \textbf{2007},
  \emph{98}, 186806\relax
\mciteBstWouldAddEndPuncttrue
\mciteSetBstMidEndSepPunct{\mcitedefaultmidpunct}
{\mcitedefaultendpunct}{\mcitedefaultseppunct}\relax
\EndOfBibitem
\bibitem[Zomer et~al.(2011)Zomer, Dash, Tombros, and van Wees]{APL-Paul2011}
Zomer,~P.~J.; Dash,~S.~P.; Tombros,~N.; van Wees,~B.~J. \emph{Applied Physics
  Letters} \textbf{2011}, \emph{99}, 232104\relax
\mciteBstWouldAddEndPuncttrue
\mciteSetBstMidEndSepPunct{\mcitedefaultmidpunct}
{\mcitedefaultendpunct}{\mcitedefaultseppunct}\relax
\EndOfBibitem
\end{mcitethebibliography}

\providecommand*\mcitethebibliography{\thebibliography}
\csname @ifundefined\endcsname{endmcitethebibliography}
  {\let\endmcitethebibliography\endthebibliography}{}

%%%%%%%%%%%%%%%%%%%%%%%%%%%%%%%%%%%%%%%%%%%%%%%%%%%%%%%%%%%%%%%%%%%%%%%%%%%%%%%%%%%%%%

\end{document}